\documentclass[a4paper,11pt]{article}
\pdfoutput=1 

\usepackage{jcappub} 

\usepackage[T1]{fontenc} 

\usepackage{placeins}
\usepackage{caption}
\usepackage{color}

\definecolor{cblue}{RGB}{100,5,255}
\definecolor{cred}{RGB}{255,50,40}

\title{\boldmath Constraints on Brane-World Inflation from the CMB Power Spectrum: Revisited}

\author[a,b, 1]{Mayukh~R.~Gangopadhyay,}\note{Corresponding author.}
\author[a]{Grant~J.~Mathews} 

\affiliation[a]{Center for Astrophysics,
Department of Physics,\\ University of Notre Dame, Notre Dame, IN 46556, USA}
\affiliation[b]{Theory Division, Saha Institute of Nuclear Physics, 1/AF, Bidhannagar, Kolkata- 64, India}
\emailAdd{mayukh.raj@saha.ac.in}
\emailAdd{gmathews@nd.edu}

\abstract{We analyze the Randal Sundrum brane-world inflation scenario in the context of  the latest CMB constraints from {\it Planck}.  We summarize constraints on the most popular classes of models and explore some more realistic inflaton effective potentials. The constraint on standard inflationary parameters changes in the brane-world scenario. We confirm that in general the brane-world scenario increases the tensor-to-scalar ratio, thus making this paradigm less consistent with the {\it Planck} constraints.  
 Indeed, when BICEP2/Keck constraints are included, all monomial potentials in the brane-world scenario become disfavored compared to the standard scenario.  However,  for natural inflation the brane-world scenario could fit the constraints better due to larger allowed values of $e$-foldings $N$ before the end of inflation in the brane-world.
}

\begin{document}
\maketitle
\flushbottom

\section{Introduction}
\label{sec:intro}
Inflation is required to solve various  problems in standard big-bang cosmology. It also provides the primordial density fluctuations required for the formation of large scale structure \cite{Liddle,cmbinflate}. A number of inflationary models have been suggested (cf. \cite{Liddle}) since the first proposed inflationary paradigm. In the last few years observations by the Wilkinson Microwave Anisotropy Probe ({\it WMAP}) \cite{WMAP9} and the {\it Planck} mission \cite{PlanckXX}  have constrained the inflationary paradigm. In particular, the {\it Planck15} inflation analysis \cite{PlanckXX} has  ruled out many of the frequently adopted models. 

The three most important parameters when constraining  inflation models are the spectral index ($n_s$), the tensor-to-scalar ratio ($r$) and the running of the spectral index ($\alpha = {dn_s}/{dlnk}$). It is now clear, for example,  that  models like the quadratic or quartic effective inflaton potentials are ruled out based upon the parameters deduced from the CMB power spectrum. Constraints on brane-world inflation have also previously been studied by many authors e.g.~\cite{PlanckXX, Maartens00, Goheer02, Bennai06, Calcagni04, Tsujikawa04, Calcagni14, Neupane14, Okada15} ( see also reviews in Refs. \cite{Langlois03, Maartens10}).  Although there are a number of braneworld paradigms, here we consider inflation 
 in the context of the Randal Sundrum brane-world scenario (RS II) \cite{Randall99,Randall99a}.   In this paper we reinvestigate  a number of inflationary potentials in the context of the newest {\it Planck15} \cite{PlanckXX}  constraints.  There have been similar recent studies based upon the {\it Planck15} \cite{Okada15} or the {\it Planck15} plus  {\it WMAP} polarization \cite{WMAP9, WP}, and BAO \cite{BAO} data. 
We also consider the  more stringent constraints based upon the combined {\it Planck15} \cite{PlanckXX} + BICEP2/Keck \cite{BICEP2} data, and   we consider several  monomial potentials that were not explicitly analyzed in previous studies.

In the brane-world scenario, the universe is a sub-manifold embedded in a higher-dimensional spacetime. Physical matter fields are confined to this sub-manifold. However gravity can reside in the higher-dimensional spacetime. This scenario was first proposed \cite{Randall99} to solve the hierarchy problem in the Standard Model of particle physics. The huge difference between the electroweak and gravity scales can be solved in this scenario because the existence of a large extra dimension brings the scale of gravity down to the weak scale. This helps to eliminate the Planck scale hierarchy with respect to the electroweak scale by generating another difference of scale between the weak scale and the size of the extra dimensions. In particular, it has been shown \cite{Randall99,Randall99a} that the  hierarchy problem is solved by introducing non-compact extra dimensions. This is an alternative to the standard Kaluza-Klein compactification scheme.  Specifically, the universe is described as a three-brane embedded in a five-dimensional anti-deSitter space $AdS_5$. This scheme guarantees the usual four dimensional spacetime in our 3-brane.

The cosmological evolution of the brane-world can be reduced \cite{Binetruy00,Maartens10} to a generalized Friedmann equation whereby the cosmological expansion for an observer in the three brane is described by:\\
\begin{equation}
\left(\frac{\dot{a}}{a}\right)^2
=\frac{8 \pi G_{\rm N}}{3} \rho
-\frac{K}{a^2}+\frac{\Lambda_{4}}{3}
+\frac{\kappa_{5}^4}{36}\rho^2 + \frac{\mu}{a^4}~~.
\label{Friedmann}
\end{equation}
Here, $a(t)$ is the usual scale factor at time $t$, while $\rho$ is the energy density of matter in the normal 3 space. $G_N$ is the four dimensional normal gravitational constant and is related 
to it's five-dimensional counterpart $\kappa_5$ by,
\begin{equation}
G_{\rm N} = \kappa_{5}^4 \lambda / 48 \pi~~,
\end{equation}
where $\lambda$ is the intrinsic tension on the brane, $\kappa_5^{2}= M_5^{-3}$, and  $M_5$ is the five-dimensional Planck mass. The $\Lambda_4$ in the third term is the four dimensional cosmological constant and is related to it's five dimensional counterpart by, 

\begin{equation}
\Lambda_{4}= \Lambda = \kappa_{5}^4 \lambda^2 /12 + 3 \Lambda_{5}/4~~.
\end{equation}
Note, that for  $\Lambda_{4}$ to be close to zero $\Lambda_{5}$ should be negative. 

The standard Friedmannian cosmology does not contain the fourth and the fifth terms of Eq. (\ref{Friedmann}). The fifth term scales like radiation with a constant
$\mu$. It is called the dark radiation. This term derives from the electric part of the five-dimensional Weyl tensor. The coefficient  $\mu$ is a constant of integration obtained by integrating the five-dimensional Einstein equations. The magnitude and sign of $\mu$ depend upon the initial conditions. The effects of the dark radiation term have been previously well studied \cite{Ichiki02,Sasanka16,sasankan17}.

In this paper, however,  we are primarily interested in the fourth term of Eq.~(\ref{Friedmann}). The fourth term arises from the imposition of a junction condition for the scale factor at the surface of the brane. The modified Friedmann equation at high energy where this term dominates, makes the early universe cosmology different from the standard scenario \cite{braxbruck}. During the post-reheating, radiation-dominated epoch this term vanishes very quickly as $a^{-8}$. Nevertheless, it can play a significant role in the inflationary era when the universe is dominated by vacuum energy. If the standard inflationary potentials are inserted as part of the $\rho^2$ term  in Eq.~(\ref{Friedmann}), there can be a significant differences between the inflation in  the projected brane-world inflation and that of  standard inflationary cosmology.  \\*

\section{Inflationary paradigm in the brane-world}
To analyze of the inflationary paradigm in the brane-world it is useful to simplify the Friedemann equation. To achieve this we first assume that the effect of the dark radiation term can
 be neglected during the very early time of the inflation epoch.  This is justified since the magnitude of this term is limited to be small   \cite{Ichiki02,Sasanka16} from the combined BBN and CMB constraints. We also take the curvature term  $K$ to be zero  during inflation along with the cosmological constant term. This leads to:
\begin{equation}
\left(\frac{\dot{a}}{a}\right)^2
=\frac{8 \pi G_{\rm N}}{3} \rho
+\frac{\kappa_{5}^4}{36}\rho^2 ~~.
\label{Friedmannmodified}
\end{equation}

Then writing the equations in terms of the reduced Planck mass $M_P$ and writing $\rho$ as the inflation generating potential we obtain:

\begin{equation}
H^2 = \frac{V(\phi)}{3M_{P}^{2}} \biggl(1+\frac{V(\phi)}{\rho_0}\biggr)~~.
\label{hubble}
\end{equation}
Here, $V(\phi)$ is the inflaton potential, $\phi$ is the usual inflaton field and the variable $\rho_0$ is defined as:
\begin{equation}
\rho_0 = 12 \frac{M_{5}^{6}}{M_{P}^{2}}~~.
\end{equation}
Thus, the usual slow roll paradigm \cite{Liddle} is altered due to this change in the evolution equation. It is worth mentioning that this scenario approaches  the standard cosmology when, $V/ \rho_{0}<<1$. In the following sections, we would carry out the calculations in the limit, $V/ \rho_{0}>>1$. Since the inflaton is confined to the brane, the scalar perturbation obeys a similar  evolution to that of the standard cosmology after taking into account  \cite{Maartens10, Calcagni14} the modification of the Hubble expansion.  Nevertheless, the gravitational power spectrum and the tensor spectrum change due to the presence of the extra dimension. This can be understood  intuitively  from the fact that the graviton resides in the extra dimension. Thus, there is an effect of the extra dimension  in the tensor perturbation.
In particular, the scalar perturbation becomes:
\begin{equation}
P_s = \frac{9}{4\pi^2}\frac{H^6}{V'^2}~~.
\end{equation}

The spectral index $n_s = d\ln{P_s}/d\ln{k}$ and the running of the spectral index $\alpha = d n_{s}/d\ln{k}$  also change due to the modified slow roll parameters derived from the modified Hubble expansion \cite{Maartens10}. The modified slow roll parameters are:
\begin{equation}
\epsilon =\frac{{ln(H^2)}'{V}'}{6H^2}~~,\eta = \frac{{V}''}{3H^2}~~.
\end{equation}
Here, the prime stands for the derivative with respect to $\phi$ and  from now on, we take, $M_{P} = 1$. 

In standard inflation analytic expressions for $n_s$, $r$,  and $\alpha$ can be derived via simple calculations \cite{Liddle}.   Extending this to  the brane-world  is straightforward \cite{Maartens00, Maartens10, Calcagni14, Okada15} allowing for the modification of the Hubble parameter Eq. (\ref{hubble}). From this, the spectral index $n_s$ and running of the spectral index $\alpha$ become \cite{Okada15},
\begin{equation}
n_s = 1 - 6 \epsilon + 2 \eta ~~,~~ \alpha = \frac{V'}{3 H^2} \bigl( 6 \epsilon' - 2 \eta' \bigr)
\label{eq:index}
\end{equation}

The most interesting change occurs for the tensor power spectra. Because gravity resides in the higher dimensional manifold,  an extra correction term is required. Hence, the tensor power spectrum becomes \cite{Langlois00}:
\begin{equation}
P_T = 8\biggl(\frac{H}{2\pi}\biggr)^2F(x_0)^2~~,
\end{equation}
where the extra factor $F(x_0)$ is written :
\begin{equation}
F(x) = \biggl(\sqrt{1+x^2}  - x^2 \ln{\biggl[\frac{1}{x} + \sqrt{1+\frac{1}{x^2}}\biggr]}\biggr)^{-1/2}~~,
\end{equation}
with, $x_0 = 2(3H^2/\rho_0)^{1/2} $.
For $x_0\ll 1$ this reduces to the standard cosmology. For ,  $x_0\gg1$ the correction factor can be approximated by $\sqrt{3x_0/2}$. 
The tensor-to-scalar ratio is then just,
$r \equiv P_T/P_s$.

Finally, the number of $e$-folds before the end of inflation that the observable scales left the horizon can be calculated in the  usual way \cite{Liddle} leading to:
\begin{equation}
N = \int_{\phi_e}^{\phi_0}d\phi \frac{3H^2}{{V}'} = \int_{\phi_e}^{\phi_0}d\phi \frac{V}{{V}'}\biggl(1+\frac{V}{\rho_0}\biggr) ~~.
\label{Neq}
\end{equation}
Here, $\phi_0$ and $\phi_e$ represent the field values at the horizon exit of the CMB modes and the end of inflation respectively. There is an independent bound \cite{Wang04} on this number of $e$-folds in the brane world of $N \leq  75$. However, that limit was calculated for a closed universe with a positive cosmological constant which does not re-collapse, while the present work is  for a flat universe. Nevertheless,, as discussed in \cite{leach}, the number of e-foldings before the end of inflation that observable scales left the horizon may be larger in the brane-world scenario than in the standard scenario. As such, for illustration we consider $N = 50, 60$ and 70 in the case of the brane-world scenario. We note, however, that a detailed calculation of the number of e-folds in the brane-world scenario is still desired and we hope to address in a future work.

\section{Monomial $V(\phi) \propto \phi^n $ Inflation Models}
Analytic expressions for $n_s$, $r$,  and $\alpha$ based upon monomial inflation in the braneworld can be found via simple calculations \cite{Maartens00, Maartens10, Calcagni14}  as described in the previous section.  A comparison of the  {\it Planck} constraints with predictions for the RSII cosmology for powers $n = 2/3, 2$ can be found in \cite{Calcagni14}.  In \cite{Okada15} $n= 2$ and  $4$ are analyzed  along with other potentials.  In this study we consider a broader range of monomial potentials $n= 2/3, 1, 4/3, 2, 3, 4$.  Unlike other recent work \cite{Calcagni14, Okada15} we also consider  the more stringent constraints from the combined {\it Planck15} + BICEP2/Keck analysis.

\subsection{\bf{$V(\phi)\propto \phi$}}
We begin our discussion  of inflation in  brane-world cosmology with the Linear Axion Monodromy potential \cite{McAllister10}.  
Wrapped branes in string compactifications introduce a monodromy that extends the field
range of individual closed-string axions to beyond the Planck scale. This
leads to a  general mechanism for chaotic inflation  based upon monodromy-extended closed-string
axions.   At leading order the effective potential  can be approximated:
\begin{equation}
V(\phi)= \lambda _1\phi~~.
\end{equation}
We studied this model both in the standard and brane-world scenarios. For the standard inflation scenario one has:
\begin{equation}
n_s = 1 - \frac{6}{1+4N} ~~, r = \frac{16}{1+4N}~~,\alpha= \frac{-24}{(1+4N)^2}~~,
\end{equation}
where $N$ is the number of $e$-folds before  the end of inflation as defined  by Eq.~(\ref{Neq}).

\begin{figure}[htb]
\includegraphics[width=6in,clip]{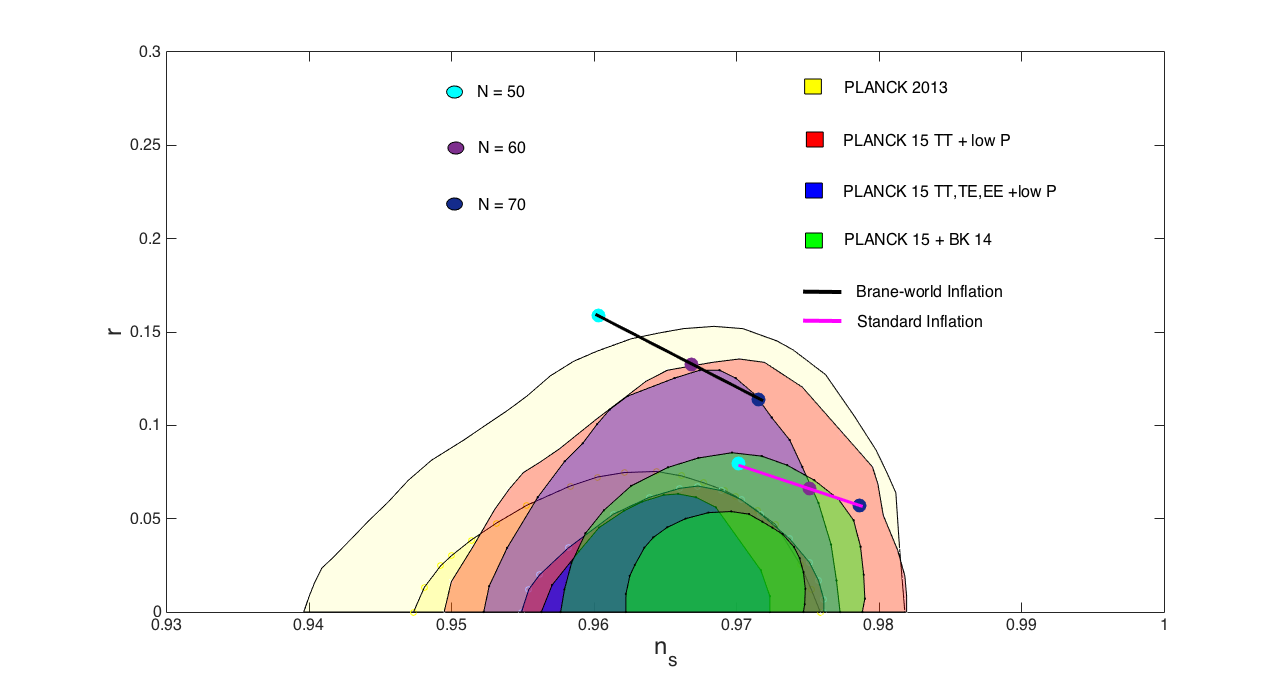} 
\caption{ Tensor-to-scalar ratio $r$ vs.~spectral index $n_s$ for the case of the Linear Axion Monodromy potential $V(\phi)= \lambda _1\phi$ in the brane-world (black line) compared to that of standard inflation  (indigo line).  Contours are the 1 and 2$\sigma$ confidence limits from various versions of the {\it Planck} analysis \cite{PlanckXX} as labeled.  }
\label{fig:1}
\end{figure}

Then taking into consideration the power spectrum measured $P_{s}(k_{0}) = 2.196 \times 10^{-9}$ (for the chosen pivot scale at $k_0$ = 0.002 Mpc$^{-1}$), we obtain,
\begin{equation}
\lambda_1[{\rm Standard}]= \frac{24\sqrt{2}\pi^2\times 2.196\times10^{-9}}{(1+4N)^{3/2}}~~({\rm GeV}^3).
\label{normal1}
\end{equation}
In the brane-world scenario, we deduce:
\begin{equation}
n_s = 1 - \frac{6}{1+3N} ~~, 
r = \frac{24}{1+3N}~~,\alpha= \frac{-18}{(1+3N)^2}~~.
\label{brane1}
\end{equation}
In this case there is a dependence of $\lambda_1$ on  $M_5$.  The analytic dependence of $\lambda_1$ is calculated to be,
\begin{equation}
\lambda_{1}[{\rm Brane-world}]= \frac{1.76\times 10^{-3}\times M_{5}^{3}}{(3N+1)}~~({\rm GeV}^3).
\label{normal1}
\end{equation}
As can be seen in Figure \ref{fig:1}, the spectral index is  reduced in the brane-world scenario.  Although the spectral index $n_{s}$,  decreases with respect to the standard cosmology, there will be always an enhancement of the tensor-to-scalar ratio in the brane-world.  This  can be seen in Equation (\ref{brane1}).
 
 \begin{figure}[!tbp]
  \centering
  \begin{minipage}[b]{0.41\textwidth}
    \includegraphics[width=\textwidth]{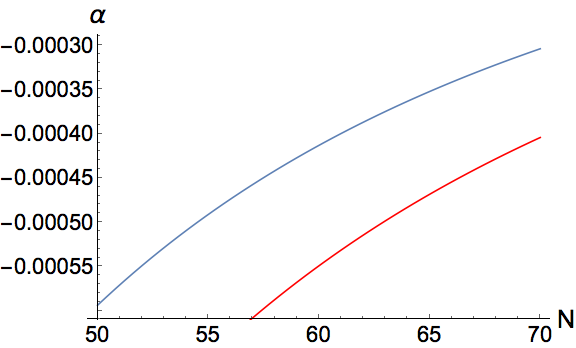}
    \caption{(Color online) The running of the spectral index as a function of  $N$ for the Linear Axion Monodromy potential in the case of brane-world inflation (red line) and standard inflation (blue line).}
 \label{fig:2}
   \end{minipage}
  \hfill
  \begin{minipage}[b]{0.57\textwidth}
     \includegraphics[width=\textwidth]{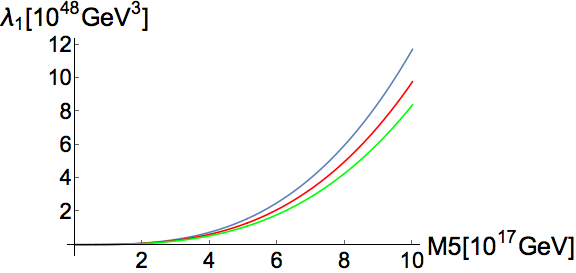}
    \caption{(Color online) The constant term $\lambda_1$ as a function of $M_5$ for various values of  $N$ during inflation: $N = 50$ (Blue) , $N = 60$ (Red) and $N = 70$ (Green).}
  \label{fig:3}
 \end{minipage}
\end{figure}

Figure  \ref{fig:2} illustrates the dependence of the running spectral index on the number of $e$-folds of inflation before the end of inflation that the observable scales left the horizon. The magnitude of the running of the spectral index $\alpha$ increases in the brane world.  Figure \ref{fig:3} shows the dependence of $\lambda_1$ on $M_5$ for various values of $N$. One can see  that the running spectral index varies with  the number of $e$-folds before the end of inflation (cf. Eq.~\ref{brane1}).  Also, values for $\lambda_1$ vary significantly within the plotted range for $M_5$.  We note that a lower bound to  $M_5$ can be  fixed by requiring $\rho_0^{1/4} > T_{BBN}$ at the epoch of big bang nucleosynthesis \cite{Ichiki02, Okada15}.  However a more stringent range for $M_5$ is determined \cite{Okada15}  by fits to the  CMB {\it Planck15} contours (c.f. Fig.~\ref{fig:1}). This constraint is described in detail in Section 5.

The effect of the brane-world scenario can even be seen on the scalar power spectrum through the running of the spectral index. The magnitude of the running increases (becomes more negative) in the braneworld scenario. Although the effect is small,  there is a discernible difference with respect to the case of standard inflation.(cf. Equation (\ref{brane1})).

\subsection{\bf{$V(\phi)\propto \phi^2$}}
 Next we investigate the popular quadratic potential:
 \begin{equation}
V(\phi)= \frac{1}{2}m^2\phi^2~~.
\end{equation}
 For the standard scenario one can write,
\begin{equation}
n_s = 1 - \frac{4}{1+2N} ~~, r = \frac{16}{1+2N}~~,\alpha= \frac{-8}{(1+2N)^2}~~,
\end{equation}
Then again taking the measured power spectrum in consideration as before, we obtain  the mass, 
\begin{equation}
m[{\rm Standard}]=\frac{2.196\times 10^{-9}\times 12 \pi^{2}}{(2N+1)^{2}}~~ ({\rm GeV}).
\label{normal1}
\end{equation}

while in the brane-world scenario we obtain,
\begin{equation}
n_s = 1 - \frac{5}{1+2N} ~~, r = \frac{24}{1+2N}~~,\alpha= \frac{-10}{(1+2N)^2}~~.
\end{equation}
The mass in the case of brane-world inflation is:
\begin{equation}
m[{\rm Brane-world}]=\frac{2.69527\times 10^{-3}M_{5}}{(N+\frac{1}{2})^\frac{5}{6}}~~ ({\rm GeV}).
\label{normal1}
\end{equation}

\begin{figure}[htb]
\includegraphics[width=6in,clip]{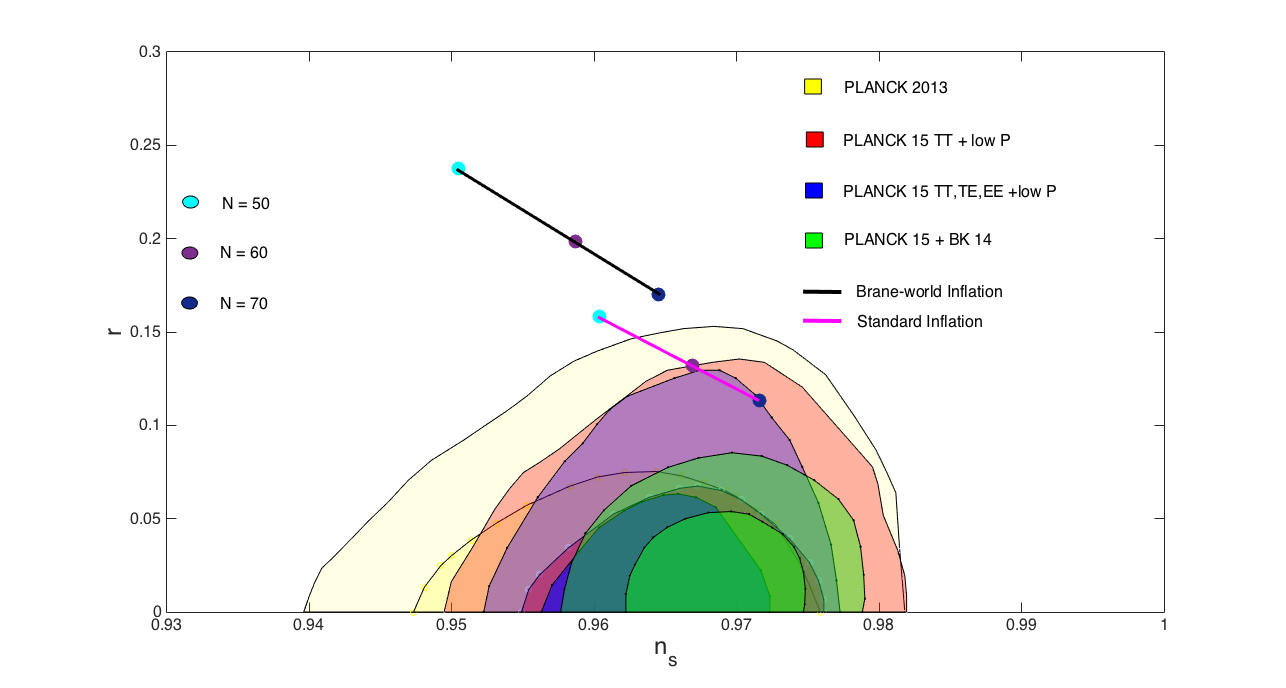} 
\caption{ Tensor-to-scalar ratio $r$ vs.~spectral index $n_s$ for the case of $V(\phi)= (1/2)m^2\phi^2$in the brane-world (black line) compared to that of standard (indigo line) inflation.  Contours are the 1 and 2$\sigma$ confidence limits from various versions of the {\it Planck} analysis \cite{PlanckXX} as labeled. }
\label{fig:4}
\end{figure}
Figure \ref{fig:4} shows the tensor-to-scalar ratio $r$ versus the spectral index $n_s$  as a function of $N$ for the case of $V(\phi)= (1/2)m^2\phi^2$ in both the brane world and  standard inflation.  These are compared to the 1 and 2$\sigma$ confidence limit contours  from the {\it Planck} analysis \cite{PlanckXX}.  Here we see again that in brane-world inflation for this potential the  
$r$ always exceeds that of standard inflation.  Hence, although the standard inflation paradigm is marginally allowed at the $2\sigma$ level in the {\it Planck TT,TE,EE+low L} analysis \cite{PlanckXX}, the brane-world inflation paradigm with this potential is ruled out at more than 2$\sigma$.

\begin{figure}[!tbp]
  \centering
  \begin{minipage}[b]{0.44\textwidth}
    \includegraphics[width=\textwidth]{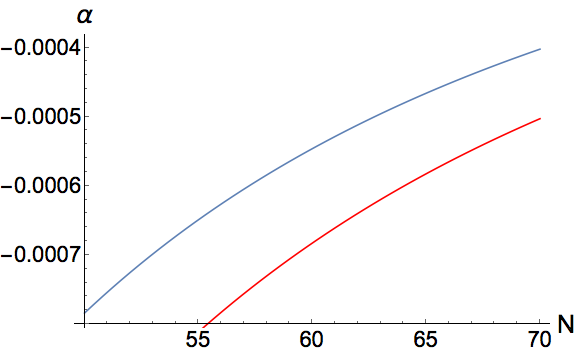}
    \caption{The running of the spectral index as a function of $N$ for the brane world (red line) and standard inflation (blue line)  in the case of $V(\phi)= (1/2)m^2\phi^2$.}
 \label{fig:5}
  \end{minipage}
  \hfill
  \begin{minipage}[b]{0.55\textwidth}
    \includegraphics[width=\textwidth]{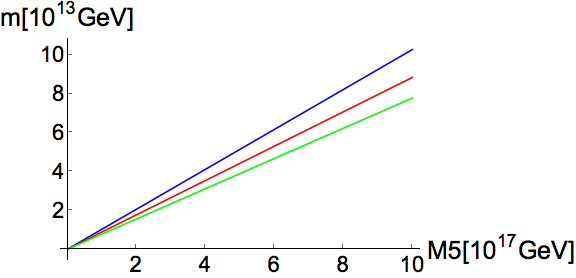}
    \caption{The mass term $m$ with as a function of  $M_5$ for  N = 50 (Blue), N = 60 (Red) and N = 70 (Green) in the case of $V(\phi)= (1/2)m^2\phi^2$.}
 \label{fig:6}
  \end{minipage}
\end{figure}

Figures \ref{fig:5}  illustrates the dependence of the running spectral index on $N$ for the quadratic effective potential.  Figure \ref{fig:6} shows the dependence of the mass $m$ on $M_5$ for various values of $N$.  One can see  that values for $m$ vary linearly within the allowed range for $M_5$.  Also, the dependence of the running of spectral index on $N$ is significant.  

\subsection{$V(\phi)\propto \phi^3$}
\begin{figure}[htb]
\includegraphics[width=6in,clip]{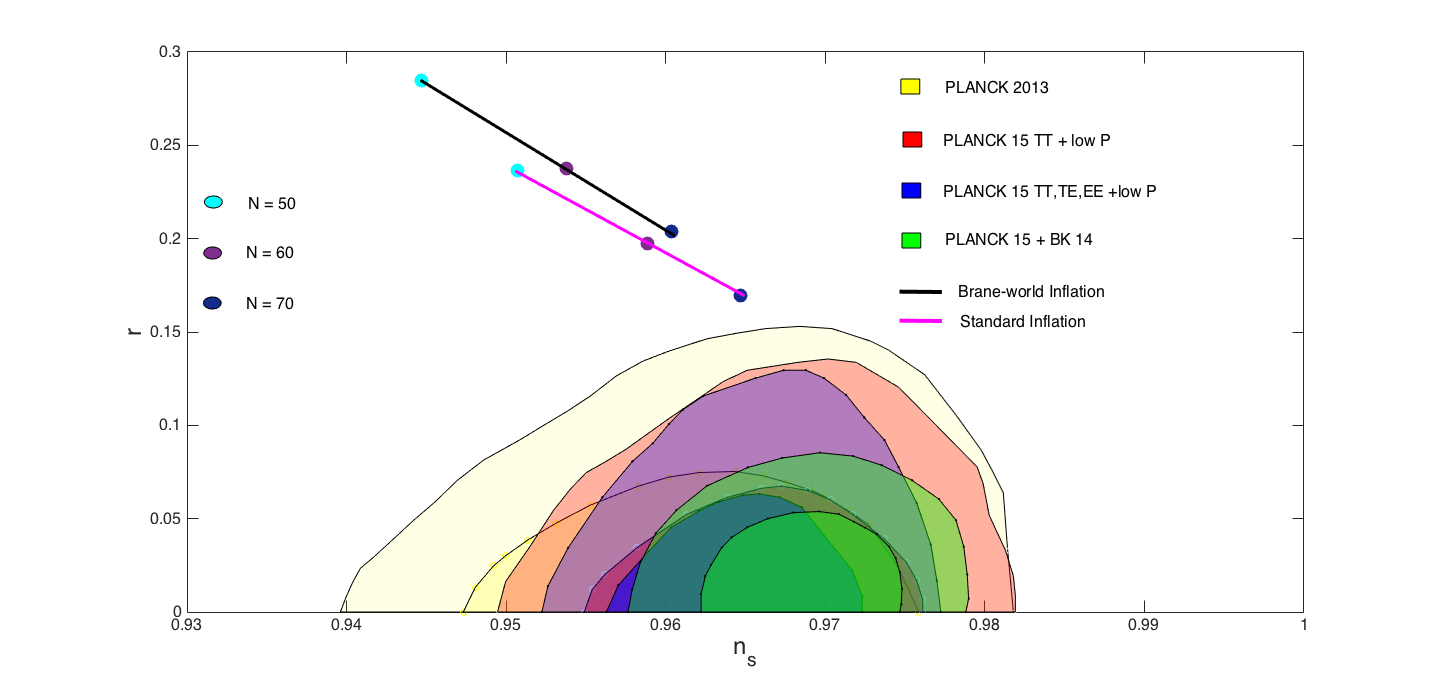} 
\caption{ Tensor-to-scalar ratio $r$ vs.~spectral index $n_s$ for the case of $V(\phi)= \lambda _3\phi^3$ in the brane-world (black line) compared to that of standard (indigo line) inflation.  Contours are the 1 and 2$\sigma$ confidence limits from various versions of the {\it Planck} analysis \cite{PlanckXX} as labeled. }
\label{fig:41}
\end{figure}
The cubic  potential has previously been shown \cite{PlanckXX} to be  well outside the allowed parameter space allowed by the {\it Planck15} analysis. Nevertheless,  for  completeness we summarize  the form of the spectral index, the tensor-to-scalar ratio, and the running of the spectral index. In the standard cosmology one  obtains,
\begin{equation}
n_s = 1 - \frac{10}{3+4N} ~~, r = \frac{48}{3+4N}~~,\alpha= \frac{-40}{(3+4N)^2}~~.
\end{equation}
Again with the same constraint from the measured power spectrum, we obtain,  
\begin{equation}
\lambda_3[{\rm Standard}]= \frac{8.4931\times10^{11}}{(3+4N)^{5/2}}~~({\rm GeV})~~.
\label{normal3}
\end{equation}

\begin{figure}[!tbp]
  \centering
  \begin{minipage}[b]{0.44\textwidth}
    \includegraphics[width=\textwidth]{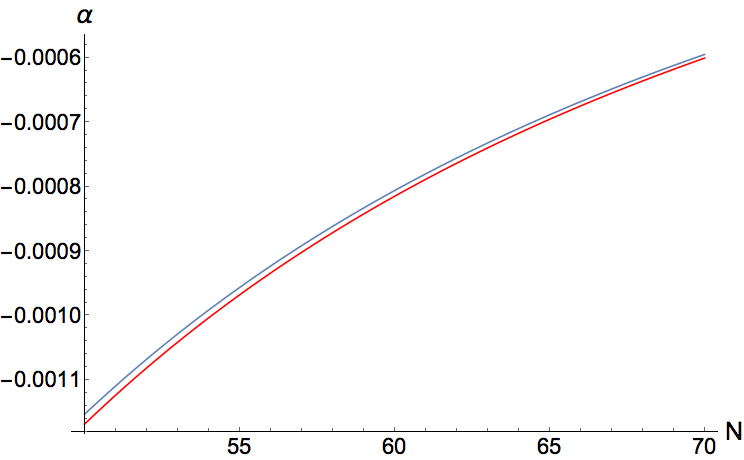}
       \caption{The running of the spectral index as a function of  $N$  for both the brane-world (blue line) and standard (red line) inflation scenarios and $V(\phi)= \lambda _{3}\phi^{3}$.}
  \label{fig:42}
 \end{minipage}
  \hfill
  \begin{minipage}[b]{0.55\textwidth}
    \includegraphics[width=\textwidth]{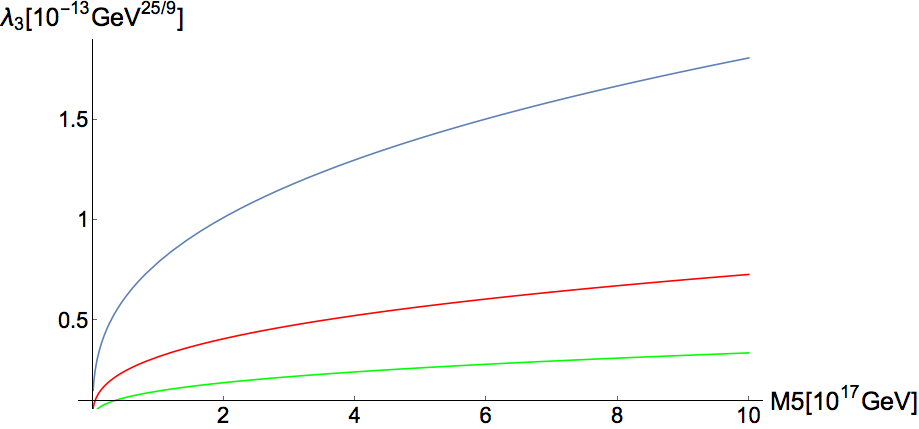}
    \caption{The constant term $\lambda_{3}$ as a function of  $M_5$ for  $N = 50$ (Blue), $N = 60$ (Red) and $N = 70$ (Green).}
   \label{fig:43}
\end{minipage}
\end{figure}

While the brane-world scenario we deduce,
\begin{equation}
n_s = 1 - \frac{14}{3+5N} ~~, r = \frac{72}{3+5N}~~,\alpha= \frac{-70}{(3+5N)^2}~~.
\end{equation}
In the same manner, we have calculated $\lambda_1$. In this case there is a dependence of $\lambda_3$ on  $M_5$.  The analytic dependence of $\lambda_3$ is calculated to be,
\begin{equation}
\lambda_{3}[{\rm Brane-world}]= 7.73846\times10^{-8}\times[\frac{M_5^{0.2}}{(5N+3)^{2.8}}]^{1.8}~~({\rm GeV}^{25/9}).
\label{normal31}
\end{equation}
Figure \ref{fig:41} shows the tensor-to-scalar ratio $r$ versus the spectral index $n_s$  as a function of $N$ for the case of cubic potential. In Figure \ref{fig:42}, the dependence of $\alpha$ on $N$ are demonstrated for the cubic potential, while   Figure \ref{fig:43} demonstrates  the dependence  of $\lambda_{3}$ on $N$ and $M_5$.

\FloatBarrier

\subsection{$V(\phi)\propto \phi^4$}
\begin{figure}[htb]
\includegraphics[width=6in,clip]{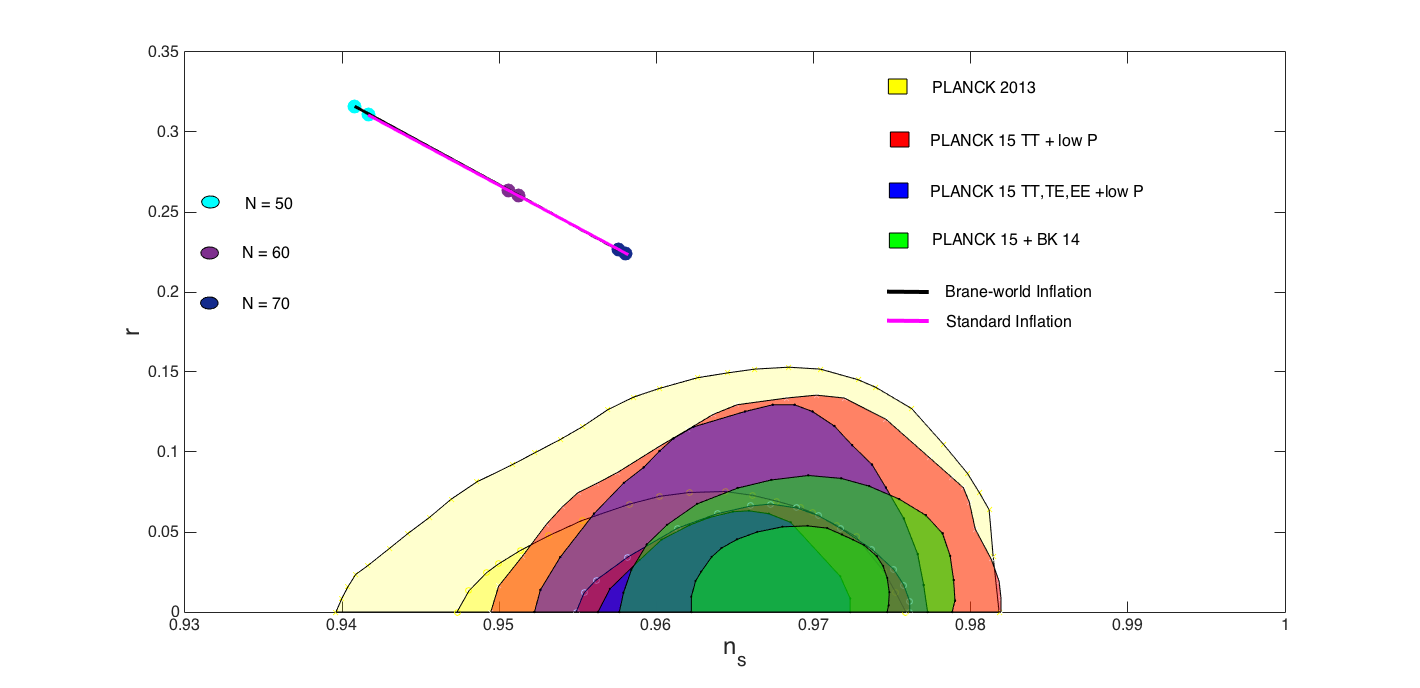} 
\caption{Tensor-to-scalar ratio $r$ vs.~spectral index $n_s$ for the case of $V(\phi)= \lambda _4\phi^4$ in the brane-world (black line) compared to that of standard (indigo line) inflation.  Contours are the 1 and 2$\sigma$ confidence limits from various versions of the {\it Planck} analysis \cite{PlanckXX} as labeled.  }
\label{fig:7}
\end{figure}

\begin{figure}[!tbp]
  \centering
    \includegraphics[width=4in,clip]{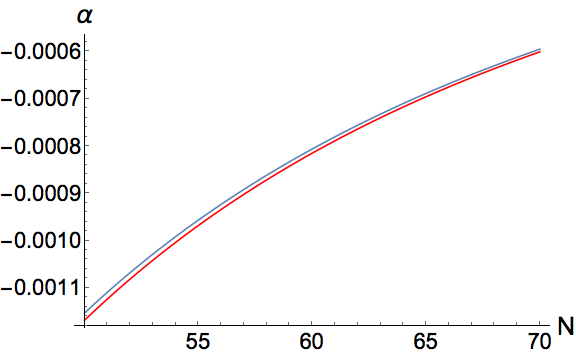}
       \caption{The running of the spectral index as a function of  $N$  for both the brane-world (blue line) and standard (red line) inflation scenarios and $V(\phi)= \lambda _{4}\phi^{4}$.}
  \label{fig:72}
\end{figure}

For the case of the quartic potential $V(\phi)= \lambda _4\phi^4$, in the standard cosmology we get,
\begin{equation}
n_s = 1 - \frac{6}{3+2N} ~~, r = \frac{32}{3+2N}~~,\alpha= \frac{-12}{(3+2N)^2}~~.
\end{equation}
For the quartic potential, we obtain,  
\begin{equation}
\lambda_4[{\rm Standard}]= \frac{8.225\times10^{-10}\pi^2}{(1+N)^{3}}~~.
\label{normal41}
\end{equation}
In the brane-world scenario we deduce,
\begin{equation}
n_s = 1 - \frac{9}{2+3N} ~~, r = \frac{48}{2+3N}~~,\alpha= \frac{-27}{(2+3N)^2}~~.
\end{equation}
Here again we have calculated $\lambda_4$. In this case there is no dependence of $\lambda_4$ on  $M_5$. This is expected because in case of the quartic potential the normalization is unit-less (thus independent of $M_5$).  This  maintains the renormalizability of the potential.  Hence, the results are very similar to the standard scenario. The analytic form of $\lambda_4$ is calculated to be,
\begin{equation}
\lambda_{4}[{\rm Brane-world}]= \frac{3\pi^2\times2.196\times10^{-9}}{8(2+3N)^3}~~.
\label{normal42}
\end{equation}
As shown in Figure \ref{fig:7} there is not much shift  in $n_s$ and $r$ between the brane-world scenario and the standard cosmology for this potential. In Figure \ref{fig:72}, the variation of $\alpha$ is plotted as a function of  $N$.
\subsection{$V(\phi)\propto \phi^{2/3}$}
The $V(\phi)= \lambda _{2/3}\phi^{2/3}$ is probably  the most realistic effective potential among the simple monomial potentials. 
This is a lowest order approximation to   axion monodromy  \cite{Silverstein08}. It  is a string-theory motivated version of natural inflation \cite{Freese90, Adams93, Freese93, Filipe05}. 
The analytic continuation on a compact manifold of this class of model makes it a more realistic candidate for the brane-world scenario.
Therefore, before analyzing the original natural inflation, we investigate this case. This effective potential  avoids the super-Planck scale width of the cosine part of the 
natural inflation potential \cite{Freese90, Adams93, Freese93, Filipe05}.  

In standard inflation with this potential one obtains,
\begin{equation}
n_s = 1 - \frac{8}{1+6N} ~~, r = \frac{16}{1+6N}~~,\alpha= \frac{-48}{(1+6N)^2}~~.
\end{equation}
Again we calculated the constant $\lambda_{\frac{2}{3}}$ in standard cosmology to obtain:
\begin{equation}
\lambda_{\frac{2}{3}}[{\rm Standard}]=\frac{2.196\times 10^{-9}\times 12\times 6^{\frac{2}{3}}\times \pi ^{2}}{(6N+1)^{\frac{4}{3}}} ~~ {\rm (GeV^{\frac{10}{3}})}.
\label{normal1}
\end{equation}

In the brane-world scenario, however,  we deduce,
\begin{equation}
n_s = 1 - \frac{7}{1+4N} ~~, r = \frac{24}{1+4N}~~,\alpha= \frac{-28}{(1+4N)^2}~~.
\end{equation}
Again we calculated the constant $\lambda_{\frac{2}{3}}$ in brane-world cosmology at the pivot scale:
\begin{equation}
\lambda_{\frac{2}{3}}[{\rm Brane-world}]=\frac{7.853\times 10^{-2}\times M_{5}^{\frac{10}{3}}}{(4N+1)^{\frac{7}{9}}} ~~ {\rm (GeV^{\frac{10}{3}})}~~.
\label{normal1}
\end{equation}

\begin{figure}[htb]
\includegraphics[width=6in,clip]{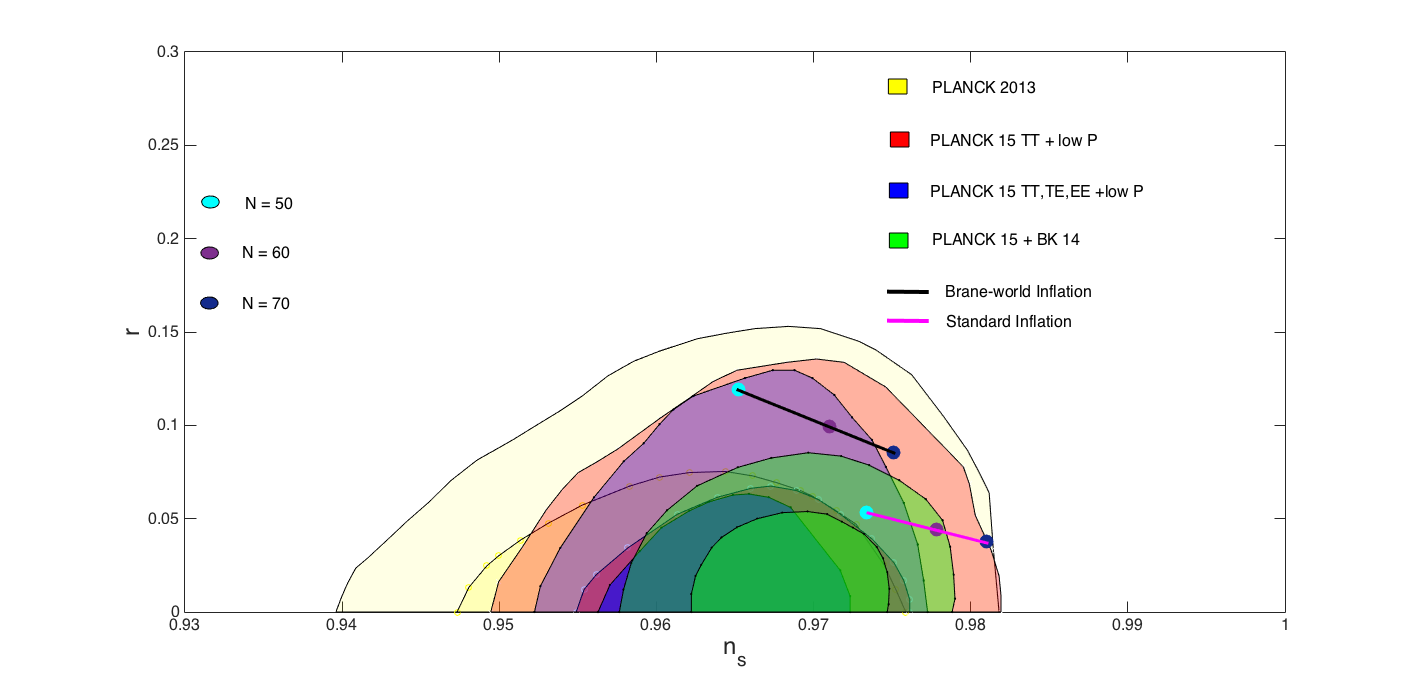} 
\caption{ Tensor-to-scalar ratio $r$ vs.~spectral index $n_s$ for the case of $V(\phi)= \lambda _{2/3}\phi^{2/3}$ in the brane-world (black line) compared to that of standard (indigo line) inflation.  Contours are the 1 and 2$\sigma$ confidence limits from various versions of the {\it Planck} analysis \cite{PlanckXX} as labeled.}
\label{fig:8}
\end{figure}

Figure \ref{fig:8} summarizes the tensor-to-scalar ratio $r$ versus $n_s$ for standard and brane-world inflation with the axion monodromy potential. 
The {\it Planck} analysis \cite{PlanckXX} tends to disfavor values for the  spectral index that are close to unity $n_s \rightarrow 1$ as is the case for standard inflation with this potential. However,  in brane-world inflation the scalar index falls near  the optimum value for $n_s$ from the {\it Planck TT + low-P} analysis, even though the value of the tensor-to-scalar ratio is increased.  The standard inflation values, however, remain a better fit when the BICEP2/Keck analysis is also included.

\begin{figure}[!tbp]
  \centering
  \begin{minipage}[b]{0.44\textwidth}
    \includegraphics[width=\textwidth]{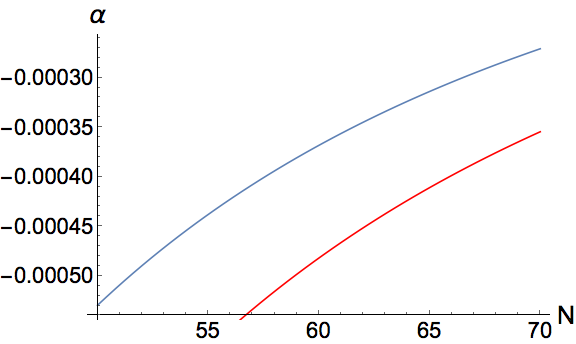}
       \caption{The running of the spectral index as a function of $N$  for both the brane-world (blue line) and standard (red line) inflation scenarios and $V(\phi)= \lambda _{2/3}\phi^{2/3}$.}
  \label{fig:9}
 \end{minipage}
  \hfill
  \begin{minipage}[b]{0.55\textwidth}
    \includegraphics[width=\textwidth]{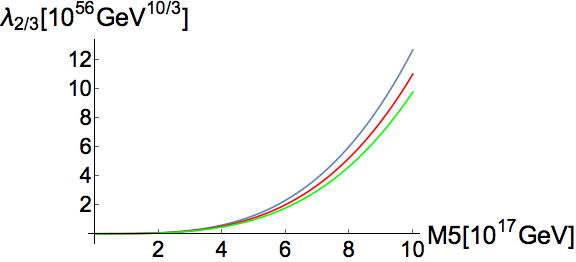}
    \caption{The constant term $\lambda_{2/3}$ as a function of  $M_5$ for  $N = 50$ (Blue), $N = 60$ (Red) and $N = 70$ (Green).}
   \label{fig:10}
\end{minipage}
\end{figure}

Figure \ref{fig:9} illustrates  how the running of the coupling constant $\alpha$ changes as a function of $N$ in the brane-world model compared  to the standard inflation. Figure  
\ref{fig:10} shows  how the constant term  $\lambda_{2/3}$ changes with  $M_5$.

\subsection{$V(\phi)\propto \phi^{4/3}$}
The $\phi^{4/3}$ potential is studied next. For the standard inflation scenario,
\begin{equation}
n_s = 1 - \frac{20}{3+12N} ~~, r = \frac{64}{3+12N}~~,\alpha= \frac{-80}{3(1+4N)^2}~~.
\end{equation}
Again we calculated the constant $\lambda_{\frac{4}{3}}$ in standard cosmology.
\begin{equation}
\lambda_{\frac{4}{3}}[{\rm Standard}]=\frac{2.196\times 10^{-9}\times 16\times 3^{\frac{2}{3}}\times \pi ^{2}}{(4N+1)^{\frac{5}{3}}} ~~ {\rm [GeV^{8/3}}.
\label{normal1}
\end{equation}

\begin{figure}[!tbp]
  \centering
  \begin{minipage}[b]{0.44\textwidth}
    \includegraphics[width=\textwidth]{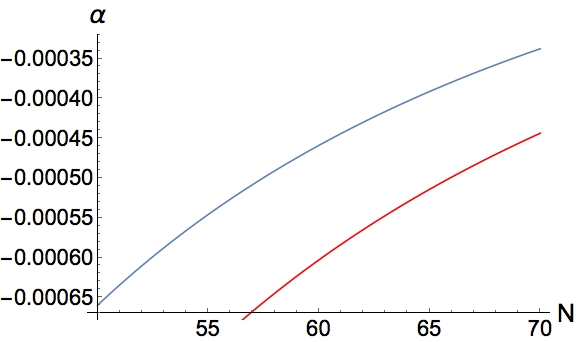}
    \caption{The running of the spectral index as a function of  $N$ for both scenarios for the case of $V(\phi)= \lambda _{4/3}\phi^{4/3}$.}
  \label{fig:12}
 \end{minipage}
  \hfill
  \begin{minipage}[b]{0.55\textwidth}
    \includegraphics[width=\textwidth]{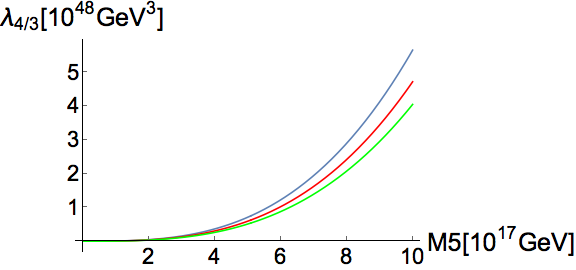}
    \caption{The constant term $\lambda_{4/3}$  as a function of  $M_5$ for $N = 50$ (Blue), $N = 60$ (Red) and $N = 70$ (Green).}
   \label{fig:13}
\end{minipage}
\end{figure}

However, for the brane-word scenario we obtain,
\begin{equation}
n_s = 1 - \frac{11}{2+5N} ~~, r = \frac{48}{2+5N}~~,\alpha= \frac{-55}{(2+5N)^2}~~.
\end{equation}
Again we calculated the constant $\lambda_{\frac{4}{3}}$ in brane-world cosmology.
\begin{equation}
\lambda_{\frac{4}{3}}[{\rm Brane-world}]=\frac{1.423\times 10^{-3}\times M_{5}^{3}}{(5N+2)}~~ {\rm [GeV^{8/3}]}.
\label{normal1}
\end{equation}
\begin{figure}[htb]
\includegraphics[width=6in,clip]{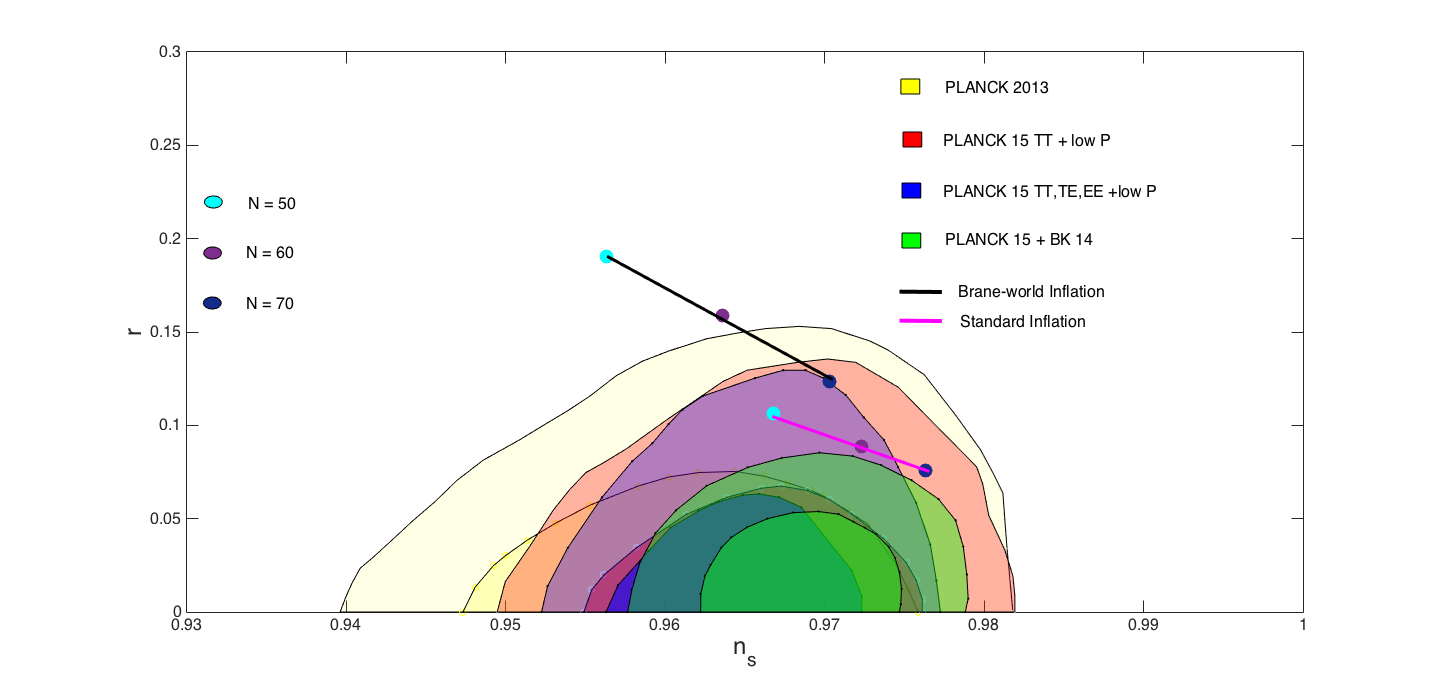} 
\caption{Tensor-to-scalar ratio $r$ vs.~spectral index $n_s$ for the case of $V(\phi)= \lambda _{4/3}\phi^{4/3}$ in the brane-world (black line) compared to that of standard (indigo line) inflation.  Contours are the 1 and 2$\sigma$ confidence limits from various versions of the {\it Planck} analysis \cite{PlanckXX} as labeled.}
\label{fig:11}
\end{figure}
In this case, the brane-world scenario is just outside the Planck 2015 TT, TE, EE + low P 2$\sigma$ limit, while the standard scenario is still allowed at 2$\sigma$ (as can be seen in Figure \ref{fig:11}). However, if we include BICEP2/Keck data, the standard scenario is also ruled out at 2$\sigma$. Figure \ref{fig:12} illustrates  how the running of the coupling constant $\alpha$ changes as a function of $N$ in RS model compared  to the standard inflation. Figure  \ref{fig:13} shows  how the constant  $\lambda_{4/3}$ changes with  $M_5$.

\section{Natural Inflation Model}
In natural inflation \cite{Freese90, Adams93, Freese93, Filipe05} the role of the inflaton is played
by a  pseudo Nambu-Goldstone boson  that appears when an approximate global symmetry is broken spontaneously. 
The flatness of the potential is protected by a shift symmetry
under $\phi \rightarrow \phi +$constant, which remains after the global symmetry is spontaneously
broken. 

The  explicit breaking of the shift symmetry leads to
a potential of the form \cite{Freese90}:
\begin{equation}
V(\phi) = \Lambda^4 [1 + \cos(\phi/f)]~~.
\end{equation}
The two mass scales $\Lambda$ and $f$ characterize the height and width of the potential, respectively. The mass scale of $f$ is typically taken as $f \sim m_{pl} \sim 10^{19}$ GeV and $\Lambda\sim M_{GUT} \sim 10^{16 }$ GeV, as expected in particle physics.
Natural inflation is a very well studied model. 
We take into account, the possibility that the number of $e$-folds before the end of inflation can be increased in the brane-world scenario. A larger value of $N$ is allowed in the brane-world scenario as discussed at the end of section 2.
The parameter   $f$ determines the curvature of the potential. For standard inflation the symmetry breaking scale is constrained to be $f > 0.9 m_{pl}$ from the Planck analysis \cite{PlanckXX}.  For the brane-world, we utilize the prior  range  on $f$ that was  adopted in \cite{PlanckXX} of $ 0.3 \le \log_{10}{(f/m_{pl})} \le 2.5$.  This is chosen to encompass a reasonable range around the Planck scale and include the {\it Planck15} allowed range as seen below. For the standard inflationary cosmology the calculation of the spectral index, tensor-to-scalar ratio and the running of the spectral index is much easier because  we know that  inflation occurs while the inflaton slowly evolves towards the potential minimum at $\phi= \pi f$. Also, for  standard cosmology  the slow roll parameters, and thus $n_s$, $r$ and $\alpha$,  are dependent upon the width of the potential  $f$,  but not on the height $\Lambda$.

   For standard inflation, approximate expressions for $n_s$, $r$ and $\alpha$  can be written as:
\begin{equation}
n_s \approx 1+\frac{1}{f^2}-\frac{2}{N}~~, r \approx \frac{8}{f^2}(\frac{f^2-N}{N})~~, \alpha\approx \frac{N-f^2}{f^2 N^2}~~.
\label{eq:4.2}\end{equation}

In the brane-world scenario the derivation of the inflation parameters is much more complex as we now outline.  In  brane-world inflation, 
as in the other cases studied here, there is a  tendency for  higher values of the  tensor-to-scalar ratio than that from  standard inflation. However,  the brane-world   calculation is a bit more cumbersome for  two reasons. For one, in standard natural the parameters are only affected by $f$ (cf. Eq.~\ref{eq:4.2}).  However, in the RSII brane-world, there in an effect from  $\Lambda$ at the end of inflation. Here also the value of $\Lambda$ is taken to be $\Lambda\sim M_{GUT} \sim 10^{16 }$ GeV. Here, we have taken $\rho_{0}^{1/4}\sim10^{10}GeV$.
Ultimately, the end of the inflation occurs approximately when:
\begin{equation}
 \phi_{end}\approx\tan^{-1}{\biggl(\frac{1}{2}(-2-y-\sqrt{y^2+8y})\biggr)}~~,
\label{eq:4.3}
\end{equation}
where
 \begin{equation}
y \equiv \frac{\rho_{0}}{f^2\Lambda^4}~~.
\label{eq:4.4}
\end{equation}
Now let us say,
 \begin{equation}
(-2-y-\sqrt{y^2+8y})= -(a+b)~~, a= 2+y~~, b = \sqrt{y^2+8y}~~.
\label{eq:4.41}
\end{equation}
Then,
\begin{equation}
 \phi_{end}\approx\tan^{-1}\frac{-(a+b)}{2}~~.
\label{eq:4.31}
\end{equation}

Eqs. \ref{eq:4.3}  and \ref{eq:4.4} show that a dependence on $\Lambda$ and $f$ cannot be avoided. Secondly, as  inflation occurs, the inflaton slowly evolves towards the minimum and starts to oscillate. Moreover,  the presence of the bulk dimension causes the end of inflation based upon the condition of slow roll violation to be dependent upon  $M_5$ through  the $\rho^2$ term.\par
\begin{figure}[htb]
\includegraphics[width=6in,clip]{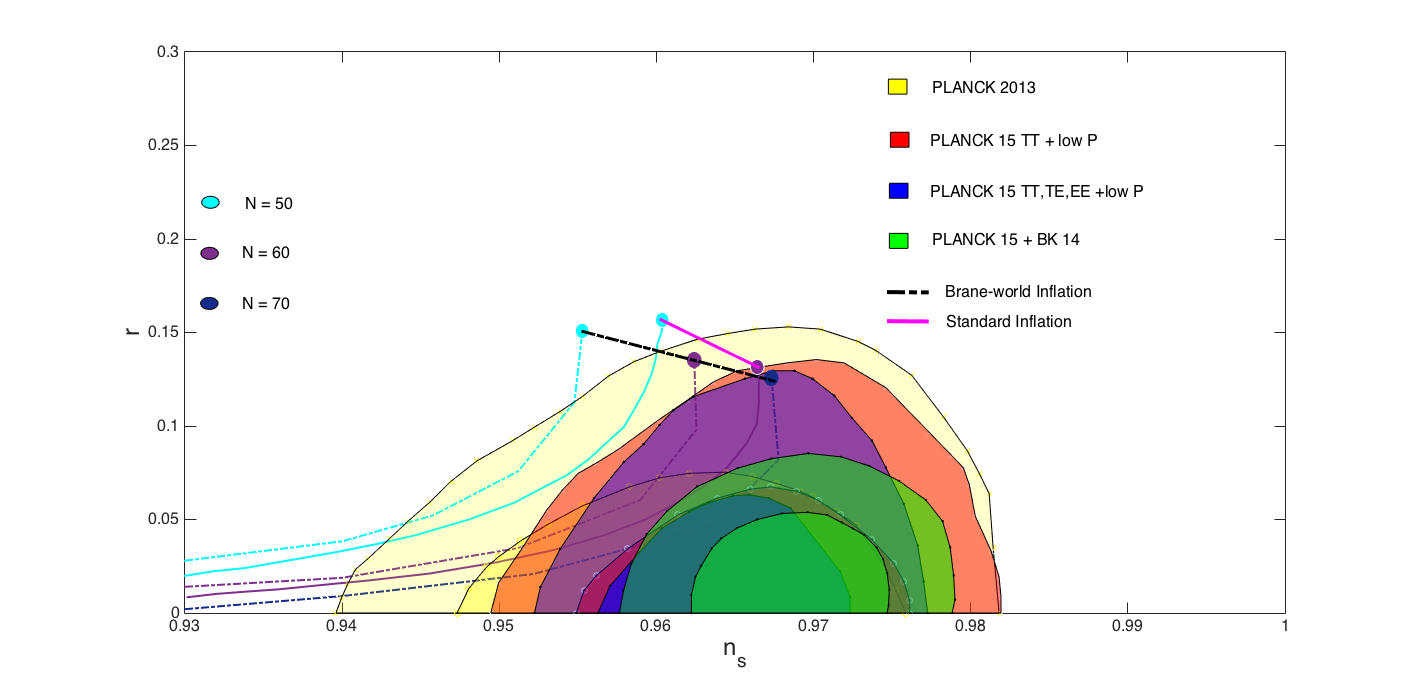} 
\caption{ Effects of brane-world cosmology in the case of natural inflation compared to  the standard inflation. The legend remains the same as in previous figures.  In this case, however,  the dashed lines denote the range of solutions for different values of $f$  in  the  brane-world scenario, while  the solid lines show  in the standard natural inflation solution.}
\label{fig:14}
\end{figure}

The expressions for $n_s$, $r$ and $\alpha$ become much more complicated in the brane-world scenario.  We have derived new relations for the spectral index, tensor-to-scalar ratio. and running of the spectral index in terms of a new auxiliary variable $z$:
\begin{equation}
n_s = 1- \frac{y}{2}\frac{z}{(1+z)^2} ~~, r = 8y\frac{-z}{(1+z)^2}~~,\alpha= 56y\frac{-z}{(1+z)^2} ~~,
\end{equation}
where,
$z$ is determined from  a solution to the exponential equation,
\begin{equation}
ze^z= d ~~,
\label{eq:4.32}
\end{equation}
and $d$ is defined as:
\begin{equation}
d= \Biggl[\frac{1}{8} \exp{\Biggl(-\biggl(1+\frac{1}{2}\bigl[(2N+1)(a-2)-b/2\bigr]\biggr)(b^2 - (a+2)b  +8) \biggr] \Biggr)}  \Biggr]^{1/2} ~~,
\label{eq:4.33}
\end{equation}
with $a$, and $b$ defined in Eq.~(\ref{eq:4.41}).  Eqs.~(\ref{eq:4.4})-(\ref{eq:4.33}) together contain an implicit  dependence of the spectral index, tensor-to-scalar ratio and the running of the spectral index on $f$ and $\Lambda$ via Eq.~(\ref{eq:4.4}).
Our solution for the brane-world natural inflation is consistent with the solution deduced in \cite{Calcagni14}. In our analysis there is a  formally different choice of auxiliary variables. We utilize the auxiliary variables $a$ and $b$ as defined above. We then numerically solve for $N$. Ref. \cite{Calcagni14} utilizes an auxiliary variable $A_{RS}$ that is related  to our parameter $y = 1/(3 A_{RS})$, and hence $a$ and $b$. In that sense the derivations in the two papers are equivalent. Values of $N$ in both approaches are then related to the adopted auxiliary variables and the value of $\cos{(\phi/f)}$ at the end of inflation. Both papers then use numerical calculations to obtain the final forms for the spectral index and other observables.\\
Hence, although appearing formally different, our calculations are equivalent  to those of \cite{Calcagni14} as noted below. 
To clarify the similarity of our result  to the result obtained in \cite{Calcagni14}, we point out that to first order the results are same.   The parameter $z$ in our calculation can be related to the parameter defined as $x$ in the section 5.2.1 in \cite{Calcagni14} as:
\begin{equation}
z\approx\frac{3 x^2-\sqrt{3} \sqrt{3 x^4+8 x^3+14 x^2+16 x+7}+4 x+5}{2 (x-1)}~~.
\end{equation}
The agreement between our results and those of Ref.~\cite{Calcagni14} then serves as a validation of both derivations. Figure \ref{fig:14} shows the tensor-to-scalar ratio $r$ versus spectral index $n_s$ for the brane-world and standard inflation scenarios. In this case the solutions are  bands as indicated by lines on the figure due to the range of values for the width parameter $f$.

From a comparison of Fig.~\ref{fig:14} with Fig.~\ref{fig:4} one can see  that in the limit of large $f$ the results coincide with those of the $\phi^2$ potential as noted in \cite{Calcagni14}.  As one can see on Fig.~\ref{fig:14}, for the same values of $N$ the standard inflation provides a better fit to the {\it Planck} constraints.  Also, our results for $N = 50$ and $60$ are nearly identical to the corresponding graph in  \cite{Calcagni14}.  However, we also point out, that because the brane-world natural inflation allows for a higher value for $N$, the region of large $f$ and $N=70$ is displaced toward higher $n_s$ and better agrees with the CMB constraints than that of the standard inflation. 

\begin{figure}[!tbp]
  \centering
  \begin{minipage}[b]{0.5\textwidth}
    \includegraphics[width=\textwidth]{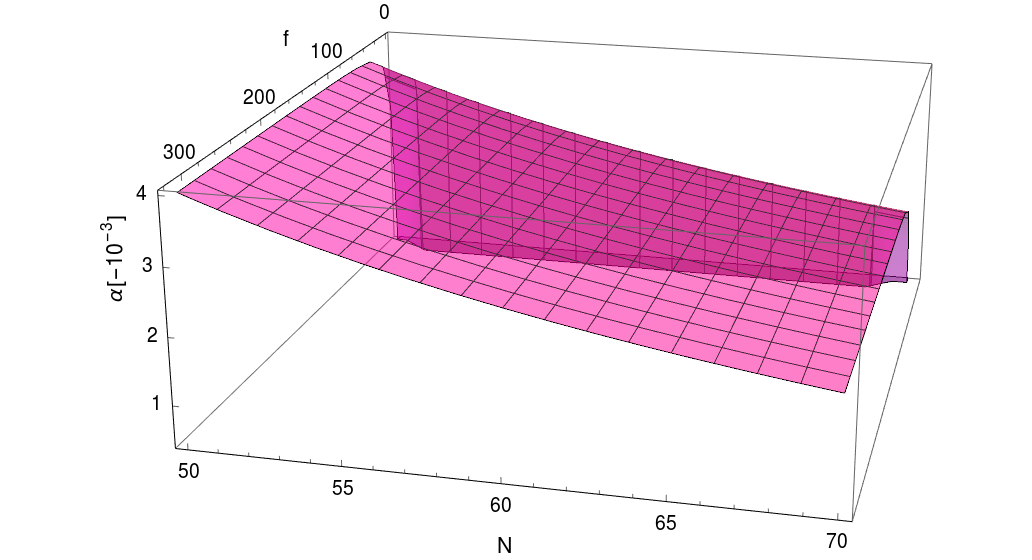}
    \caption{The running of the spectral index with respect to  $N$ and $f$ in the standard natural inflation scenario. Here $f$ is in Planck units. }
 \label{fig:15}
  \end{minipage}
  \hfill
  \begin{minipage}[b]{0.45\textwidth}
    \includegraphics[width=\textwidth]{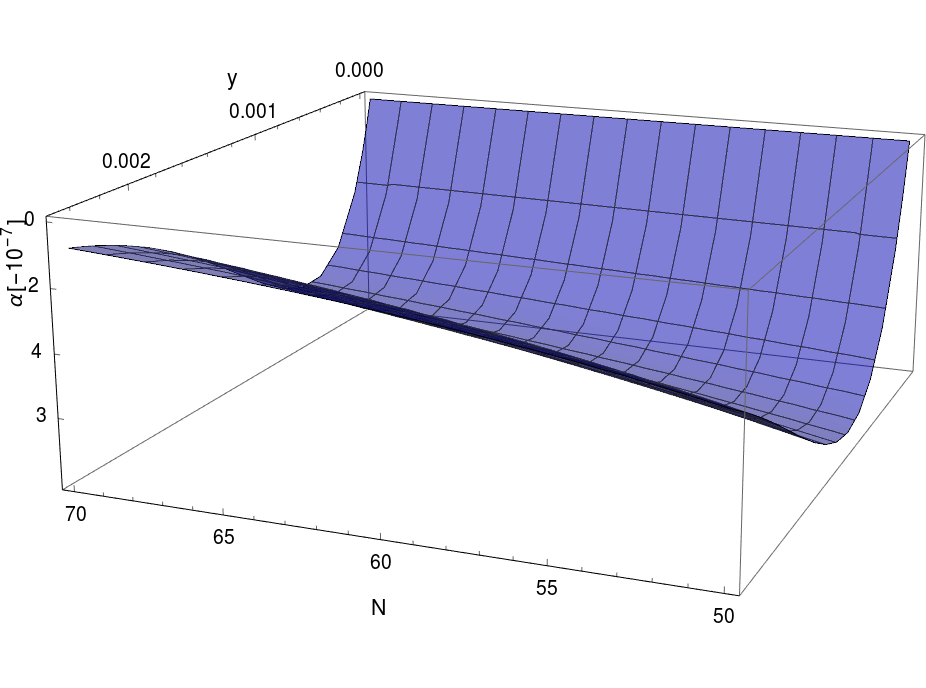}
    \caption{The running of the spectral index with respect to  $N$ and $y$ in the natural inflation scenario for  brane-world cosmology.}
 \label{fig:16}
  \end{minipage}
\end{figure}

The running of the spectral index $\alpha$  depends explicitly  on $N$ and $y$, while $y$ is related to the model parameters $f$ and $\Lambda$. Thus, it is worthwhile  to graphically illustrate  the resultant dependence of $\alpha$ upon  the various model parameters as shown in Figs.~\ref{fig:15} and \ref{fig:16}.  The running of the spectral index for the standard natural inflation as a function of $N$ and  $f$   can be seen as a surface in  three dimensions in Figure \ref{fig:15}. For the brane-world scenario, however, the dependence becomes more complicated. The running of the spectral index is not  only a function of $N$ and $f$ but also the value of $\Lambda$. In Figure \ref{fig:16} we show the dependence of $\alpha$ on N and $y$ for $N = 50-70$.

\FloatBarrier

\section{Constraints on  $M_5$}
\begin{figure}[!tbp]
  \centering
  \begin{minipage}[b]{0.45\textwidth}
    \includegraphics[width=\textwidth]{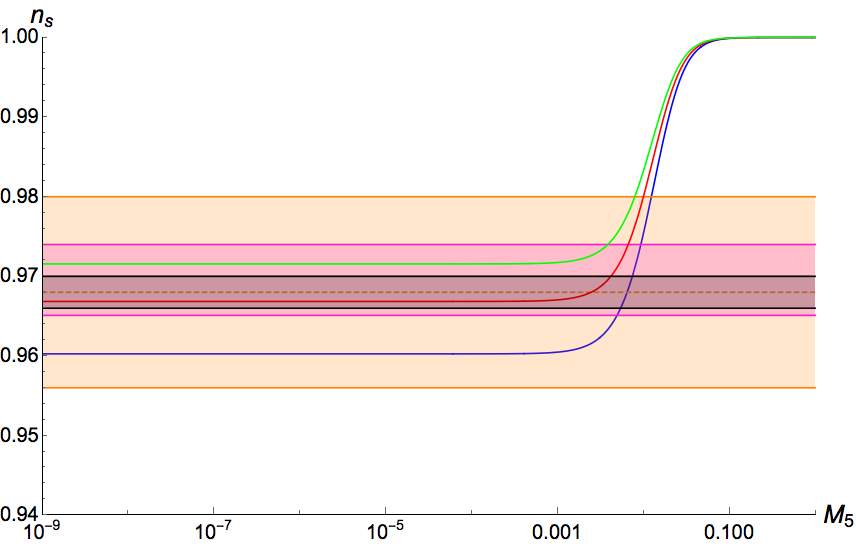}
    \captionsetup{labelformat=empty}
    \caption{(a) $V(\phi) \propto \phi $ } 
 \label{fig:22}
  \end{minipage}
  \hfill
  \begin{minipage}[b]{0.45\textwidth}
    \includegraphics[width=\textwidth]{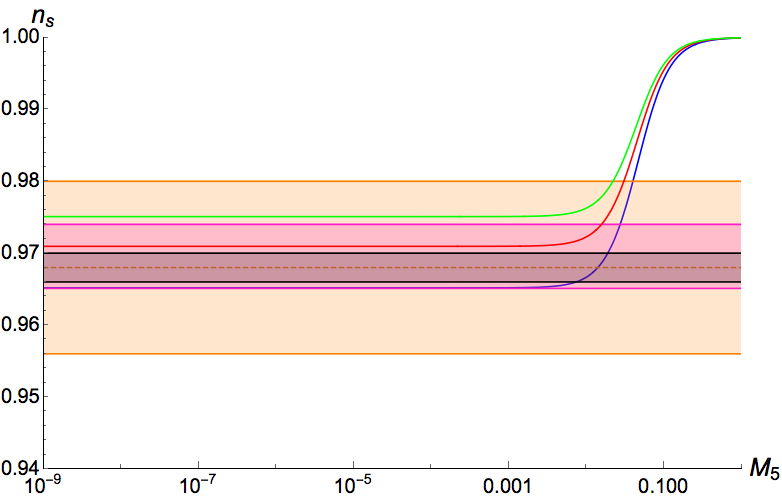}
     \captionsetup{labelformat=empty}
    \caption{(b) $V(\phi) \propto \phi^{2/3}$}
  \end{minipage}
  \hfill
  \begin{minipage}[b]{0.45\textwidth}
    \includegraphics[width=\textwidth]{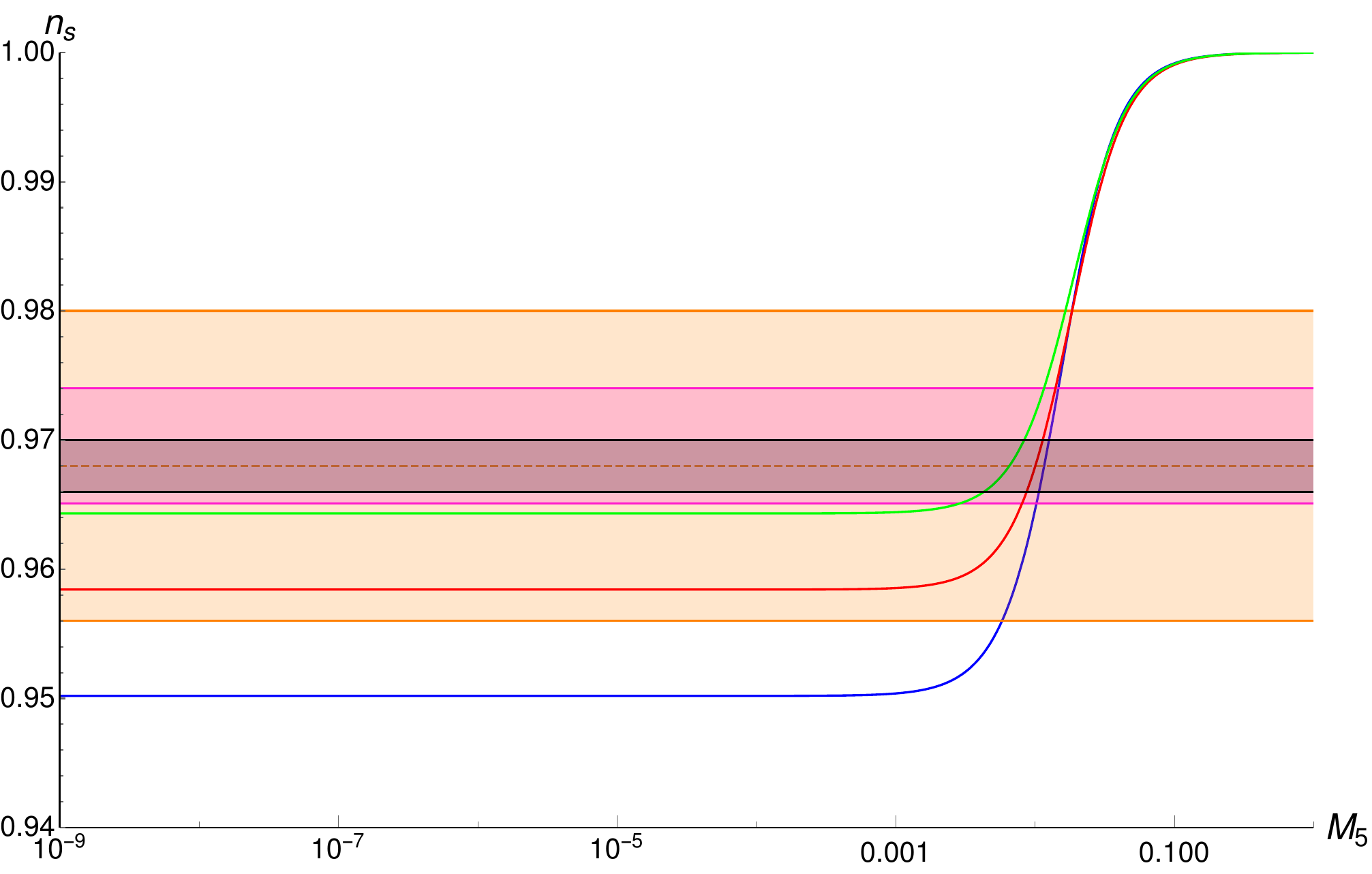}
     \captionsetup{labelformat=empty}
    \caption{(c) $V(\phi) \propto \phi^{2}$}
  \end{minipage}
 
  \hfill
  \begin{minipage}[b]{0.45\textwidth}
    \includegraphics[width=\textwidth]{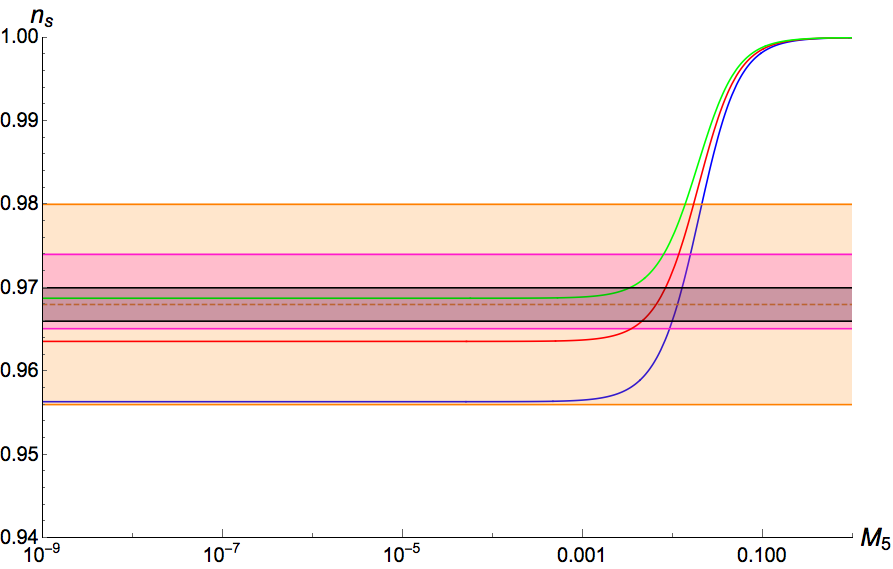}
     \captionsetup{labelformat=empty}
    \caption{(d)  $V(\phi) \propto \phi^{4/3}$}
 \end{minipage}
  \hfill
  \begin{minipage}[b]{0.45\textwidth}
    \includegraphics[width=\textwidth]{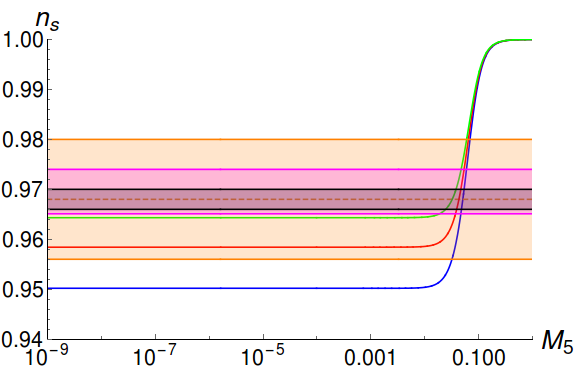}
     \captionsetup{labelformat=empty}
    \caption{(e) $V(\phi) \propto 1 + \cos{( \phi/f)}$.}
  \end{minipage}
   \captionsetup{labelformat=empty}
  \caption{\textbf{Figure 21:} Spectral index $n_s$ as a function of $M_5$ (in units of $M_p$) for different values of $N$. The blue, red and green lines correspond to $N = 50, 60, 70$ respectively.}
 \end{figure}

There is a  model-independent lower bound on the 5-dimensional Planck mass of $M_5 > 8.8$ TeV from Big Bang Nucleosynthesis (BBN) \cite{Okada15} and the requirement that $\rho_0^{1/4} \ge T_{BBN}$. Of particular relevance to the present work is that  precision measurements by {\it Planck }\cite{PlanckXX}  also allow for  constraints on $M_5$ for each of the inflation models
considered here.  This follows \cite{Okada15} from  the requirement that the spectral index fall within the 2$\sigma $ limits from the {\it Planck} analysis. In Figures \ref{fig:22}a-e, we  show the values  the spectral index $n_s$ as a function of $M_5$ for some of the models considered in the present work and various values of $N$.  The corresponding constraints on $M_5$ are summarized in Table \ref{tab:1}. It is worth mentioning that in this section there is no initial assumption of $V/\rho_o >>0$ as in the previous sections. This is to show the variation with respect to $\rho_0$. For most cases there is only an upper limit to $M_5$.  However, for the special case of $\phi^2$ and natural inflation with $N = 50$ a lower limit can also be deduced. A similar kind of analysis is carried out in \cite{Okada15} for some of the inflationary potentials. Though the only common model with ours is the $\phi^2$ potential, it should be mentioned here that the variation in that case is carried out for a set of fixed value of $n_s$. Whereas in this work  $n_s$ is varied freely and then for an allowed range of $n_s$, the range of $M_5$ is estimated and represented in a log-linear plot and $M_5$ is expressed in terms of Planck unit. Also in our analysis all the bounds on $M_5$ are kept in mind.\\

 For all the cases on Figures \ref{fig:22}a-e, the horizontal dashed brown line shows the {\it Planck} central value ($n_s = 0.968$) for the TT+lowP+lensing data \cite{PlanckXX}. The pink band corresponds to the 1-$\sigma$ region ($\Delta n_s \sim 0.006$) and the orange band corresponds to 2-$\sigma$ region. The black band corresponds to a hypothetical  future 1-$\sigma$ sensitivity if $\Delta n_s$ could ever be reduced to  $\Delta n_s \sim 0.002$; assuming the central value remains unchanged \cite{abzajian,finelli}.

\begin{table}[tbp]
\centering
\begin{tabular}{|l|r|c|}
\hline
$~~~~~~~V(\phi)$~~~&~~~~~N~~~~~& $~~~~M_5$($M_p =1$)~~~~\\
\hline \hline
$~~~~~~\propto \phi $&~~~~~50~~~~~& $M_5 < 0.0128$\\
 &~~~~~60~~~~~& $M_5 < 0.0098$ \\
  &~~~~~70~~~~~& $M_5 < 0.0079$ \\
\hline
$~~~~~~\propto \phi^{2/3}$ &~~~~~50~~~~~& $M_5 < 0.0411$ \\
   &~~~~~60~~~~~& $M_5 < 0.0315$ \\
   &~~~~~70~~~~~& $M_5 < 0.0411$ \\
\hline
$~~~~~~\propto \phi^{2}$ &~~~~~50~~~~~& $0.0059<M_5 < 0.0196$ \\
   &~~~~~60~~~~~& $M_5 < 0.0196$ \\
   &~~~~~70~~~~~& $M_5 < 0.0170$ \\
\hline
$~~~~~~\propto \phi^{4/3}$ &~~~~~50~~~~~& $M_5 < 0.0211$ \\
   &~~~~~60~~~~~& $M_5 < 0.0177$ \\
  &~~~~~70~~~~~& $M_5 < 0.0211$ \\
\hline
$\propto 1 + \cos{( \phi/f)}$ &~~~~~50~~~~~& $0.0314 < M_5 < 0.075$\\
   &~~~~~60~~~~~& $M_5 < 0.0679$\\
   &~~~~~70~~~~~& $M_5 < 0.0658$\\
\hline
\end{tabular}
\caption{\label{tab:i} Constraints on $M_5$ for various inflation effective potentials.}
\label{tab:1}
\end{table}

\section{Conclusion}

We have considered constraints on the brane-world inflation paradigm from the results of the {\it Planck15} analysis \cite{PlanckXX}.
We also consider the  more stringent constraints based upon the combined {\it Planck15} \cite{PlanckXX} + BICEP2/Keck \cite{BICEP2}, and   we analyze a few monomial potentials that were not explicitly considered in previous studies (e.g. \cite{Calcagni14, Okada15}).
We have confirmed previous 
analytic derivations in brane-world inflation for the spectral index, the tensor-to-scalar ratio, and the running of the spectral index for a variety of monomial inflation effective potentials  
and deduced constraints on natural inflation consistent with the previous results of \cite{Calcagni14}. We also deduce new limits on the five-dimensional Planck mass $M_5$ from the requirement that the spectral index reside within the $2 \sigma$ limits deduced by the {\it Planck} analysis.

 We confirm that in general the brane-world scenario increases the tensor-to-scalar ratio thus making this paradigm less consistent with the {\it Planck} constraints.   However, in the case of the lowest order $\phi^{2/3}$ axion monodromy (if one only compares to the {\it Planck15}  TT +  low -P constraint) there might  be a slight improvement over the standard model in the case of large  $N$.  However this improvement vanishes when comparing to the combined constraint from BICEP2/Keck.
 
 Standard  natural inflation fits the {\it Planck} constraints better for the same values of $N$.  However,  due to the fact that larger values of $N$ are allowed in the brane-world paradigm,  the brane-world natural inflation could  provide a better fit to the CMB spectrum. These results are  encouraging since the axion monodromy and natural inflation  are  most consistent with an interpretation of brane-world cosmology as an approximation to string theory.

\acknowledgments

Work at the University of Notre Dame is supported
by the U.S. Department of Energy under 
Nuclear Theory Grant DE-FG02-95-ER40934. MRG wants to thank A.Banerjee, S. Bhattacharyya and N. Kumar for enlightening discussions. We would also like to thank the anonymous referee for invaluable suggestions to make this work much better.


\end{document}